\documentclass[10pt]{article}
\usepackage{latexsym}
\usepackage{graphicx}
\usepackage{epsfig}
\usepackage{url}
\usepackage{amsmath,amssymb}
\usepackage{xspace}
\usepackage{authblk}
\usepackage{algorithmic}
\usepackage[ruled,vlined]{algorithm2e}
\usepackage{float}

\usepackage{soul}
\usepackage{multirow}
\usepackage{color, soul}
\usepackage{comment}
\usepackage{enumitem}
\usepackage{array,tabularx}
\setlist[enumerate]{%
wide =0.6\parindent}%

\usepackage{framed,enumitem}

\usepackage[colorinlistoftodos]{todonotes}

\newenvironment{conditions*}
  {\par\vspace{\abovedisplayskip}\noindent
   \tabularx{\columnwidth}{>{$}l<{$} @{\ : } >{\raggedright\arraybackslash}X}}
  {\endtabularx\par\vspace{\belowdisplayskip}}

\title{A Novel IoT Trust Model Leveraging Fully Distributed Behavioral Fingerprinting and Secure Delegation}

\author{Marco Arazzi$^{1}$, Serena Nicolazzo$^{2}$, Antonino Nocera$^{1}$\\
\small{
$^{1}$ Department of Electrical, Computer and Biomedical Engineering, University of Pavia \\
via A. Ferrata, 5 - 27100 - Pavia, Italy
\\
$^{2}$ Department of Computer Science, University of Milan \\
Via G. Celoria, 18 - 20133 - Milan, Italy\\
marco.arazzi01@universitadipavia.it; serena.nicolazzo@unimi.it; antonino.nocera@unipv.it}
}

\date{}
\begin{document}

\maketitle
\begin{abstract}
With the number of connected smart devices expected to constantly grow in the next years, Internet of Things (IoT) solutions are experimenting a booming demand to make data collection and processing easier. The ability of IoT 
appliances to provide pervasive and better support to everyday tasks, in most cases transparently to humans, is also achieved through the high degree of autonomy of such devices. However, the higher the number of new capabilities and services provided in an autonomous way, the wider the attack surface that exposes users to data hacking and lost. In this scenario, many critical challenges arise also because IoT devices have heterogeneous computational capabilities (i.e., in the same network there might be simple sensors/actuators as well as more complex and smart nodes). In this paper, we try to provide a contribution in this setting, tackling the non-trivial issues of equipping smart things with a strategy to evaluate, also through their neighbors, the trustworthiness of an object in the network before interacting with it. To do so, we design a novel and fully distributed trust model exploiting devices' behavioral fingerprints, a distributed consensus mechanism and the Blockchain technology. Beyond the detailed description of our framework, we also illustrate the security model associated
with it and the tests carried out to evaluate its correctness and performance.

\end{abstract}

\section{Introduction}

Nowadays, the Internet of Things (IoT) has grown rapidly, attracting not only researchers but also people from industrial and commercial environments. 
Indeed, this paradigm, characterized by heterogeneous and connected devices sharing data and providing services, creates huge opportunities in numerous domains.
Radio Frequency identification (RFID), wireless sensors and
other smart technologies are integrated into a variety of
applications to create networks with enhanced capabilities in terms of sensing information about the environment and
collecting measurements or operational data from their devices. 

One of the peculiarities of this new scenario is that there is no need of constant human interventions for the {\em things} to handle data, process them, exchange messages in the network or execute instructions \cite{elkhodr2016internet}. 
A typical case, in which the user
takes almost no active role and fully relies on devices
and services to act on her/his behalf, is the smart home environment. 
Think, for instance, to sensors able to recognize the presence of humans in home rooms to switch on/off lights. The information collected and used to perform this task is one of the the most sensitive and personal (i.e., people movements inside their own homes). Hence, solving the trade-off between the functionalities provided (as well as the degree of autonomy of the objects) and the sensitivity of data exchanged to obtain them is a demanding challenge to be faced. Moreover, the need for privacy 
increases when an untrusted (or also honest-but-curious) party is involved, and, in the aforementioned scenario, data is not completely processed inside the user's object directly, but rather it is handled by some external cloud-based service. 

Other domains in which lots of sensitive information are exchanged involve the healthcare domain \cite{kodali2015implementation,nicolazzo2020privacy}. In this case, wearable devices collect data about users' health condition continuously to monitor them for medical security reasons. Also in this case, data loss or compromise can result in serious damages \cite{chacko2018security}. Moreover, if an IoT device is implanted in a human body for the good functioning of some organ, attacker gaining access to it can endanger the life of the victim.
Unfortunately, the hacking of a device is not a remote possibility, indeed, a research study carried out
by Packard, reported that more than $70\%$ of the existing IoT systems have severe
vulnerabilities due to a number of reasons, such as: insecure Web interfaces, insufficient authorization mechanisms, lack of encryption for transport, and inadequate software protection \cite{HP2015,alvarez2021risks}. 

To make some reference to recent events, think for instance to famous botnet attacks launched against a large number of IoT
devices, such as: the Mirai attack in $2016$, that infected around $2.5$ million devices connected to
the Internet through a distributed denial of service (DDoS)
attack, and the Hajime attack, which brings more sophistication to some of the techniques used by Mirai \cite{eustis2019mirai,edwards2016hajime}.

Another domain gaining benefits from the growth of IoT is the Cyber Physical System (CPS) context. In such systems, physical and software components are deeply interconnected to continuously monitor the environment, and thanks to the support of intelligent systems, CPSs are also able to make decisions based on the physical changes of the surroundings. Since CPSs control assets of critical importance (e.g., industrial control systems, power grids, transportation), neglecting security standards can lead to serious consequences.

Besides autonomy, a factor that makes the management of IoT devices even more critical is their intrinsic heterogeneity. Their different computational capabilities, memory and provided functionalities, cause that, generally, poorly secured objects can connect to the Internet and can interact with more powerful devices. Hence, these objects can represent multiple points of failure that expose the entire IoT network to possible attacks, increasing the need to define and adopt non-standard security methods \cite{hassija2019survey}. 

In this scenario, the classical countermeasures to face privacy and security threats have to be rethought taking into account
the many restrictions and limitations of IoT devices (in terms of components
and devices, computational and power resources) and even their
heterogeneous and distributed nature. Since IoT technologies and applications are so
intimately associated with people, a step forward in this direction could make consumers less reluctant to adopt this new paradigm. 

Recently, researchers are starting to exploit the possibility for nodes to collaborate to make IoT networks more robust to attacks \cite{corradini2022two,rafferty2018intelligent}. 
One area of investigation in this context is trust and reputation management, which is crucial to allow efficient collaboration among the actors of the network that might not have sufficient prior knowledge about each other \cite{liu2010trust}. Trust in IoT is a comprehensive concept that takes people, devices, and their
connections into account. It can be defined as the expectation that a thing is reliable, in other words, that it acts without harming the user or other objects in the network, is resilient to attacks, and belongs to a user who is always who he claims to be \cite{leister2012ideas,aldowah2021trust}.
To make an example, assessing that a smartwatch is trusted could mean that it always gives the right (i.e., correct, complete and fresh) information to its user when queried (for instance, time, temperature, heartbeat, and so forth). Moreover, by doing so, it should not reveal more information about user habits than necessary, and it should not act as a malicious node performing tasks it is not authorized for.

As for thing-to-thing trust evaluation, classical approaches
exploit remote attestation. Through this security service an object can evaluate the current state of a potentially hacked or compromised remote device before contacting it. Remote attestation algorithms span from heavy-weight secure hardware-based techniques (such as cryptography), to light-weight software-based ones (e.g., control-flow integrity) \cite{abera2016things}. 

More recent approaches are based on the computation of device {\em fingerprint} \cite{hamad2019iot,kostas2021iotdevid}. Fingerprint represents a set of features useful to identify
an object not relying on its classical network identities, (such as IP or MAC addresses), but exploiting the information from the packets that the device exchanges over
the network.  
Always in this context, a step forward is represented by the notion of {\em behavioural fingerprint}. All the approaches based on this concept build a device profile describing what it usually does (how it interacts with the environment) and some patterns of communication (both with other objects and humans). This profile is
used to analyze the current behavior of an object and to assess whether it is congruent with the expected one \cite{ferretti2021h2o,aramini2022enhanced,hamad2019iot,miettinen2017iot,aneja2018iot}.

However, classical trust models are usually too computational heavy and do not exploit the peculiarity of IoT nodes that can collaborate to obtain consensus. Moreover, most of the approaches based on behavioural fingerprint are centralized and, as a consequence, do not take into consideration features obtainable by analyzing message payloads.

We start from the above considerations to design a solution based on a fully distributed behavioral fingerprint computation used as an input to a general IoT
trust model. Essentially, to assess the
trustworthiness of an object in the network our approach proceeds through two steps. The former is the construction of behavioral models representing
the expected conduct of every node in the network, and the latter is a suitable monitoring activity to detected possible variations in it. 

To reach our goal, we exploit all the peculiarities of IoT nodes, such as: their being autonomous, heterogeneous, and collaborative. Indeed, in our approach, we enforce that:

\begin{itemize}
    \item More objects can learn models to represent the expected behavior of a target object unobtrusively without the human intervention (autonomy).
    \item  The training and inference phases can be obtained through a privacy-preserving collaborative delegation approach in which simpler objects are supported by more powerful nodes to implement the solution (heterogeneity).
    \item Nodes collaborate to constantly monitor other objects conduct by signaling an anomaly in their normal behaviour through a distributed algorithm based on the consensus of their neighbors (collaboration).
\end{itemize}

Another point of strength of our solution is that is it completely distributed. Indeed, to reach a fully decentralized approach, we use smart contract and Blockchain technologies to: \emph{(i)} keep trace of the
evaluation of the behavior of objects at a global level and \emph{(ii)} identify the best peers to contact
to enable the aforementioned collaborative approach. As a matter of fact,
Blockchain smart contracts are already being used to manage, control and secure IoT devices \cite{khan2018iot,christidis2016blockchains}, and, in addition, lightweight adaptations of a Blockchain have been designed to support resource-constrained IoT devices  \cite{corradini2022two}.

In the following of this paper, through a deep experimental campaign, carried out leveraging real-life smart object data, we prove that our approach is feasible and equips the nodes of an IoT network with the possibility to detect if another peer is compromised before contacting it.
Interestingly, our strategy is based on a lightweight behavioral fingerprint model suitable for IoT devices and our secure delegation strategy produces advantages also in terms of running time.

Our paper starts from the research direction described in \cite{ferretti2021h2o}, where the framework H2O (Human to Object) is presented. Node belonging to H2O can continuously authenticate
an entity in the network, providing a reliability assessment mechanism based on behavioral
fingerprinting. Our proposal extends that contribution by proposing several critical enhancements and designing novel security mechanisms to improve object interaction in IoT. 
In particular, borrowing ideas from the recent scientific literature \cite{aramini2022enhanced,Thien2019}, we detail and implement a deep
learning model for the computation of the distributed behavioral fingerprinting by making it lighter, also through a Tiny Machine Learning approach. Moreover, we propose an improved distributed consensus mechanism and design a novel secure delegation strategy to compute object's reliability.
Finally, we add a Blockchain-based solution to trace, in a
fully distributed fashion, the evolution of the behavior of objects when interacting with each other.
Through a detailed security analysis, we show that our proposal is robust and addresses its objectives also in presence of attacks. 

The outline of this paper is as follows. In Section \ref{sec:RelatedWork}, we examine the literature related to our approach. In Section \ref{sec:Description}, we give a general overview of our reference IoT model and illustrate the proposed framework in detail. In Section \ref{sec:SecurityModel}, we describe our security model. In Section \ref{sec:Experiments}, we present the set of experiments carried out to test our approach and analyse its performance. Finally, in Section \ref{sec:Conclusion}, we draw our conclusions and have a look at possible future developments of our research efforts.

\section{Related Works}
\label{sec:RelatedWork}

Since IoT environment is widely distributed and
dependable on user sensitive data, the concept of trust management is becoming a crucial prerequisite for the design of new applications in this field\cite{aldowah2021trust}.
In scientific literature, different trust definitions have been settled but, speaking about IoT devices, research community agrees to define trust as the probability that the intended behavior of a thing and its actual behavior are equivalent, given fixed
context, environment, and time \cite{voas2018internet}. 

In the last years, different approaches for trust management of IoT object-to-object communications have been provided. Typically, the mechanism through which
a node can check the current state of a potentially compromised remote device,
before interacting with it, is referred as {\em remote attestation}.
This security service is implemented by \cite{tan2017mtra}, 
where a Multiple-Tier Remote Attestation protocol, called MTRA, is presented. In particular, this framework provides two methods to monitor IoT devices on the basis of their characteristics. Specifically, less
smart devices are verified through a lighter software-based attestation algorithm, whereas more powerful ones are monitored by means of trusted hardware called Trusted Platform Module (TPM). 
Another scheme in this context, providing enhanced functionalities is proposed in \cite{kuang2019esdra}, where the authors describe a many-to-one attestation approach for device swarms. Through some redundancy, this solution reduces the possibility of single point of failure typical of architectures in which a verifier node has to assess the reliability of more IoT devices. 

Possible strategies to empower objects with a mean to trust peers in their network are provided by cryptographic techniques \cite{KaSaWa04,Chen*19,Ganeriwal*04}. Nevertheless, in an IoT scenario, key
management represents an issue due to resource constrained devices and lack of a unique standard. Moreover, besides the fact that these approaches are computational demanding, they are also not robust against internal malicious nodes having the valid cryptographic keys. Finally, another weakness is that they usually rely on an external level for the computation of node trust score.

A wide group of works focuses on the issue of IoT device identification and authentication to assess reliability \cite{bezawada2021behavioral,hamad2019iot,miettinen2017iot,aneja2018iot,kostas2021iotdevid,ursino2020humanizing}.
They start from the consideration that network identifiers like IP addresses, MAC addresses, ports
numbers, etc. have been used for identifying devices, but they can be spoofed easily. 

Hence, some of them explore the concept of {\ em device fingerprinting} as a way to identify
an object not through its classical network ids, but exploiting the information contained in the communication packets exchanged over the network. 
In particular, the work presented in \cite{hamad2019iot} analyzes a sequence of packets from high-level network traffic to extract a set of unique flow-based features. From these features a fingerprint for each device is created through machine learning techniques. In the same context, the authors of \cite{aneja2018iot} exploit the potentialities of deep learning approaches to compute a set of features useful to provide a device fingerprint.
Whereas, in \cite{miettinen2017iot} a framework called IoT Sentinel is described. This schema is able to automatically provide an anomaly detection task, identifying vulnerable devices being connected to an IoT network and enforcing mitigation measures for them. In this way, it can minimize damage resulting from their forgery. Also the proposal presented in \cite{kostas2021iotdevid} presents an IoT device identification method that models the behavior of the network packets exchanged during communication by the objects.
Some of these approaches are based on timing analysis and are designed to fingerprint
specific devices that exhibit a certain behavior, e.g., probe scans for an access point.

A step forward in this context is carried out by some approaches dealing with {\em behavioral fingerprinting}. This type of techniques focuses on more application-level features to
model objects' traits, instead of relying only on the physical and link layers, as done by models using device fingerprint.
Among these characteristics there
are: protocols, request-response sequences and any periodicity in specific typology of packets along with their sizes \cite{Thangavelu*18, aramini2022enhanced,ursino2020humanizing,ferretti2021h2o}
In particular, in \cite{Thangavelu*18}, the authors describe a distributed solution for behavioral fingerprinting in IoT exploiting a decentralized approach.
They identify some network nodes, called gateways, that can monitor objects using trained classification models, thus assuring a more salable to the solution. 
Some controller nodes, instead, are in charge of performing the training part of the solution. The feature vector identified contains $111$ dimensions.
Instead, the approach proposed in \cite{ursino2020humanizing} is related to object reliability in a Multiple Internet of Things (MIoT), defining, only theoretically, also the concept of object profile. Like our approach also this scheme is based on a consensus mechanism, but the main difference is that the reliability score is simply proportional to: \emph{(i)} the fraction of successful transactions performed by the instances, and \emph{(ii)} the reliability of the corresponding objects.

As stated in the Introduction, our work starts from the considerations analysed in \cite{ferretti2021h2o}. In \cite{ferretti2021h2o} a framework called H2O (Human to Object) is presented. 
Also nodes belonging to H2O are equipped with a mechanism
to estimate the reliability of their contacts, but there are substantial differences respect to our approach, that are worth to be detailed in the following. 

The first improvement deals with the behavioral fingerprint technique, allowing an object to assess if another one (which it usually interacts with)
has been hacked or corrupted. In H2O framework it leverages state-of-the-art approaches, whereas, in the present paper, we implement a custom model. This new model, through some technical improvements and the use of a tiny machine learning approach, makes our solution suitable for devices with limited capabilities with only a maximum increase of $1\%$ of the accuracy (see Sections \ref{sec:Fingerprint} and \ref{sec:Experiments} for all the details).

In H2O, a node can attest the reliability of another peer if multiple confirmations of the peer trust score from neighbors exist (i.e., neighbors hold a fingerprint model of the peer) and if the mean of these scores is higher than a given threshold. In the present algorithm, instead, we improve the computation of the reliability score using of Blockchain technology. Indeed, as an additional functionality, our approach also estimates the quality of the contributions of each node involved
in the reliability estimation, to make our algorithm more robust to malicious false negative scores. This value affects the reliability score provided by that node and, to be globally accessible, it is stored in a Blockchain.

Another functionality provided by this paper is the design of a strategy for secure delegation to allow less capable device to participate and benefit from the approach. This algorithm is community oriented and privacy preserving, instead in H2O framework this possibility is only mentioned without developing a detailed implementation.

As for the work presented in \cite{aramini2022enhanced}, it presents en enhanced behavioral fingerprinting
models, which considers also features derived from the analysis of packet
payloads (for instance, different types of devices and their
traffic characteristics). In our scenario, we are considering a more general IoT context in which also legacy devices are available. Therefore, starting from the two solutions above, we tried to lighten the architecture as much as possible so that it could also be used by devices characterized by medium-to-low computational power and limited functionalities.
Moreover, we reduce to the minimum possible the complexity of the machine learning model in such a way to directly involve the maximum possible number of nodes (see Section \ref{sec:Fingerprint} for details).

Another new and promising technology for establishing trust in IoT networks in a distributed way is Blockchain. Indeed, different proposals have been recently developed in order to provide forms of trust or authentication in a IoT network through this new technology \cite{DiPietro*18,hammi2018bubbles,dedeoglu2019trust}. In particular, the work presented in \cite{DiPietro*18} deals with an Obligation Chain containing obligations generated by a number of nodes, called Service Consumers. These transactions are first locally accepted by Service Providers and, then, shared to the rest of the network. This kind of framework is based on the concept of {\em Islands of Trust}, defined as portion of the IoT network where trust is managed by both a full local PKI and a Certification Authority. Also the approach in \cite{hammi2018bubbles} relies on secure virtual zones (called bubbles) where things can identify and trust each other. These bubbles are obtained through Blockchain technology.
Although Blockchain provides decentralized security and privacy, it has some drawbacks in terms of delay, energy, and computational overhead generated, that are not always suitable for most limited IoT devices.
Both the work presented \cite{dedeoglu2019trust,corradini2022two} try to overcome these limitations proposing a light architecture
for improving the end-to-end trust. The proposal presented in \cite{dedeoglu2019trust} is based on some gateways calculating the trust for sensor observations based on: {\em(i)} the data they receive
from neighboring sensor nodes, {\em(ii)} the reputation of the sensor node, and {\em(iii)} the observation confidence. If the neighboring sensor nodes are associated with different gateway
nodes, then, the gateway nodes may share the evidence
with their neighboring gateway nodes to calculate the
observation trust values. This architecture is not fully distributed and secure delegation is not performed, indeed, more powerful nodes are used as gateway.
Whereas the work presented in \cite{corradini2022two} proposes a two-tier Blockchain framework to increase the security and autonomy of smart objects in the IoT by implementing a trust-based protection mechanism. This work deals with the concept of communities of objects and relies on first-tier Blockchain that is used only to record probing transactions performed to evaluate the trust of an object in another one of the same community or of a different community. Periodically, these transactions are aggregated and the obtained values are stored in the second-tier Blockchain to be globally accessed by all the communities. 
In our approach Blockchain is solely used to keep trace of the
evaluation of the behavior of objects for the anomaly detection task and to identify the best peers to contact
to enable the aforementioned collaborative approach.

A different perspective to build a trust and reputation scheme is taken by \cite{nitti2013trustworthiness}, in which the authors investigate the
trustworthiness management in a Social Internet of Thing (SIot). A SIoT, first introduced by \cite{atzori2012social}, models device interaction as social ties, allowing an object to crawl the network for finding other (possibly heterogeneous) objects in an autonomous way and establishing friendship relationships. In  \cite{nitti2013trustworthiness} the authors combine a subjective model and an objective one. In the former, each node computes the trustworthiness of its neighbors on the basis of its own experience and on the opinion of the friends in common with it. In the latter, the information about each node is distributed and stored in a distributed hash table structure, so that this information is accessible by all the nodes in the network. 

In the following, we summarize the comparison with all the works
introduced above based on the different functionalities provided by our approach, namely:

\begin{itemize}
    \item Trust. A functionality that allows nodes in the network to assign a trust score to another node according to its behaviour.
    \item Reputation. A functionality that allows a node in the network to compute a reliability score according to its neighbors' opinion about another node, even if they have not been in contact before.
    \item Light Fingerprint. A functionality that allows the computation of behavioural fingerprint for a node suitable for a IoT scenario, in which nodes have limited capabilities.
    \item Secure Delegation. A functionality according to which some computation can be entrusted to more capable devices in a privacy preserving way.
\end{itemize}

\noindent With the letter `x' we denote that the corresponding property is provided by the cited paper.

\begin{table}[ht]
	
\centering\footnotesize
\begin{tabular}{||l|l|c|c|c|c||}
\hline
Approach	& Approach Type	& Trust & Reputation& Lightweight & Secure\\
& & & &  Scheme & Delegation\\
\hline \hline
 Our approach & Fingerprint, Consensus, Delegation & x & x & x& x\\
 
  \cite{ferretti2021h2o,ursino2020humanizing} & Fingerprint, Consensus & x & x & - & -\\
  \cite{kuang2019esdra,tan2017mtra}   & Remote Attestation  & x & - & - & -\\
  \cite{Chen*19,KaSaWa04,Ganeriwal*04}  & Cryptographic & x & - & -& -\\
  \cite{hamad2019iot,aneja2018iot,miettinen2017iot,kostas2021iotdevid,bezawada2021behavioral} & Fingerprint & x & - & -& -\\
\cite{DiPietro*18,dedeoglu2019trust,hammi2018bubbles} & Blockchain & x & x & x & -\\
\cite{nitti2013trustworthiness} & Social Network &  x & x & -& -\\

\hline
	\end{tabular}
\caption{Comparison of our approach with related ones.\label{tab:comparison}}
\end{table}

\section{Description of Our Approach}
\label{sec:Description}

\subsection{General Overview}
\label{sub:GeneralOverview}

In this section, we present a general overview of our approach. 
As stated in the Introduction, our proposal focuses on the definition of a fully distributed trust model for IoT using behavioral fingerprints.
Behavioral fingerprinting is a technique largely investigated in the scientific literature (see Section \ref{sec:RelatedWork}) to model the expected and typical conduct of an online device (typically, an IoT object) when interacting with other entities in the observed ecosystem.
In traditional behavioral fingerprinting schemes, the modeling is usually performed by a centralized super-entity that can monitor and supervise objects' interactions (e.g., a network hub, an access point, or a base station) \cite{nguyen2019diot}.
However, with explicit reference to the most recent trend of IoT, in which objects are more and more autonomous and, hence, equipped with higher computational capacities, our approach defines a solution for fully distributed behavioral fingerprinting that can, hence, be used as an input to a general IoT trust model.

According to the recent scientific literature, a behavioral fingerprint can be built by training a deep learning model on the information derived by the observed object (for which the fingerprint must be derived) and its communications.
From a computational point of view, we can distinguish between two phases: {\em (i)} the training phase, and {\em (ii)} the inference phase.
The former represents the most computational expensive one and, depending on the amount of available data and the complexity of the involved model, it may require the exploitation of medium to high computationally capable devices.
The inference phase, instead, is still a computational demanding task but far less impacting when compared to the training phase. In general, it requires low to medium computationally capable devices also considering that optimizations can be applied during the training phase to obtain lighter models for the inference \cite{han2015deep}.
It is worth notice that, in a modern IoT scenario, we can identify three categories of objects:

\begin{itemize}
    \item Basic Device ({\tt BD}, for short): low power device with limited computational power.  
    \item Capable Device ({\tt CD}, for short): devices with sufficient computational power, but for which the training phase of the machine learning model would require too much time, or would negatively impact on battery consumption and therefore a conservative use of resources is preferable.
    \item Powerful Device ({\tt PD}, for short): powerful devices with sufficient computational power and stability to both train and run machine learning models.
\end{itemize}

In general, a modern IoT is composed of a constellation of heterogeneous devices belonging to the three categories above. 
Therefore, for starters, to develop a fully distributed behavioral fingerprinting approach it is necessary to suitably orchestrate a collaboration mechanism to involve all the nodes in the solution.
Moreover, as stated above, the goal of our approach is to design a fully distributed trust model of IoT leveraging behavioral fingerprinting as fundamental component of an anomaly detection strategy. 
Therefore, information about the trustworthiness of an object in the network requires both the construction of behavioral models representing its expected conduct, and suitable monitoring activities to detected possible variations in it.
By leveraging a collaborative and community-based point of view, as already done by several works in the scientific literature \cite{corradini2022two, ferretti2021h2o, white2018iotpredict, zhang2021trusted}, in our approach we enforce that more objects can learn models to represent the expected behavior of a target one and that they can collaborate to monitoring its future conduct by signaling an anomaly in case of an unexpected and impacting variation.
Also the training and inference phases can be obtained through a privacy-preserving collaborative delegation approach in which {\tt PD} and {\tt CD} devices cooperate and provide support to {\tt BD} ones to implement the solution.

To favor interactions at a global IoT level, our approach leverages a Blockchain-based solution to support the development of the distributed trust model.
In particular, a Blockchain is used to both keep trace of the evaluation of the behavior of objects for the anomaly detection task and to identify the best peers to contact to enable the aforementioned collaborative approach. 

In practice, our solution is composed of three main components, namely: {\em (i)} a distributed behavioral fingerprinting strategy; {\em (ii)} a community-oriented secure delegation; {\em (ii)} a distributed consensus mechanism to estimate object reliability.

\begin{figure}[ht]
	\centerline{
        \includegraphics[scale=0.4]{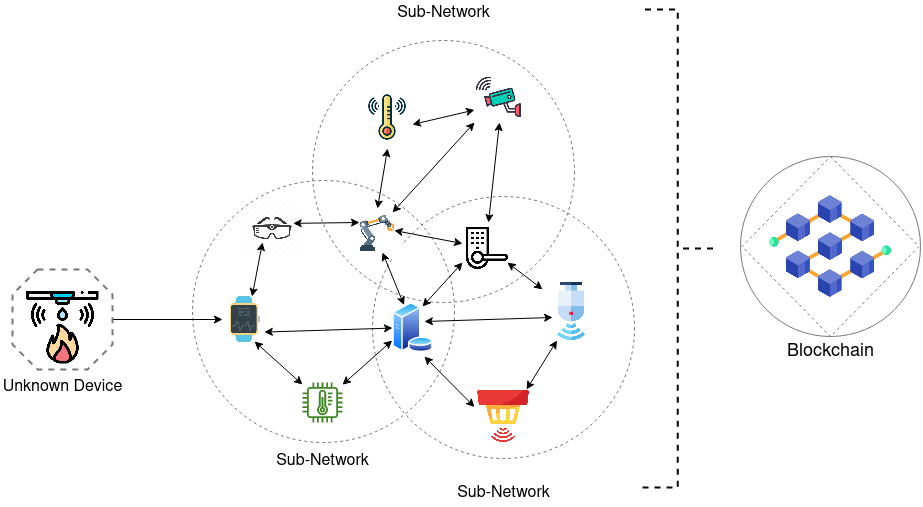}
    }
    \caption{The general architecture of our solution \label{fig:Architecture}}
\end{figure}

A general architecture of our solution is reported in Figure \ref{fig:Architecture}.

\subsection{The proposed model}
\label{sec:Model}

In this section, we introduce the model adopted to represent the main components and concepts exploited in our solution.

In particular, as explained above, we are considering a scenario characterized by the presence of different type of devices.
In our model, we consider the following concepts:
\begin{itemize}
    \item {\em The IoT node}. It represents the main actor of our system and is associated with a profile allowing the interaction with other nodes (also referred as devices or objects).
    The considered profile consists of an IoT identifier and a Blockchain account. Moreover, a node includes also all the information necessary to enable the communication with other nodes (such as the MAC address, the IP address, and so forth).
    A device $d_x$ can belong to one of the categories described in Section \ref{sec:Description}. Therefore, we can identify the following sets:
    
    \begin{itemize}
        \item $BD$: the set of basic, low power devices.
        \item $CD$: the set of limited devices that have sufficient computation power for the inference, i.e., to use a trained machine learning model, but with limitations (e.g., battery constraints or power stability) preventing them to perform a full training.
        \item $PD$: the set of powerful devices that can train machine learning models.
    \end{itemize}
    
    Finally, we define the set of all devices $D=BD \cup CD \cup PD$ of our IoT scenario.
    
    \item {\em An IoT interaction}. It represents any data exchange between a pair of devices. In particular, an IoT interaction is established whenever a generic device, $d_x$, sends a packet, $p_k$, to another device, $d_y$.
    Subsequently, we define an $interaction\_sequence$ at the time $t$, namely $is^t_{\langle d_x, d_y \rangle} = \{p_1, p_2, \dots, p_n\}$, as a consecutive packet exchange from $d_x$ to $d_y$ established with a first packet $p_1$ sent at the time $t$, such that the inter-arrival time between the packages is lower then a threshold $\tau$.
    In our representation, we preserve the direction of an $interaction\_sequence$, meaning that $is^t_{\langle d_x, d_y \rangle} \neq is^t_{\langle d_y, d_x \rangle}$
    
    Finally, we define as $communication\_set$, $C_{\langle d_x, d_y \rangle}=\{is^t_{\langle d_x, d_y \rangle}\  \forall \ t\}$, all the $interaction\_sequences$, performed at any time $t$, between $d_x$ and $d_y$.

    \item {\em The fingerprint model}. It is a behavioral fingerprint of a device $d_y$ as perceived by another device $d_x$, we define such model as:
    $$
        f_{d_x,d_y}=\mathbf{ML}(C_{\langle d_x, d_y \rangle})
    $$
    
    In words, a fingerprint model is a machine learning model built by using the $communication\_set$ between a pair of nodes as training data. The rationale is that by learning the ``original'' interaction style (i.e., the way a node sends packets to another one, as described by a suitable set of features, in the absence of anomalies), it could be possible to detect any anomalous behavior by analyzing possible variations in such a style.
    
    \item {\em The Blockchain}. It is the shared ledger used to record information about trust relationships among the IoT nodes. Our solution leverages a combination of behavioral fingerprinting and a community-oriented secure delegation strategy to enable safe interaction in IoT. For this reason, we will introduce a trust model (see Section \ref{sec:Consensus}) based on a consensus mechanism, to estimate the reliability of a node. In this solution, the Blockchain is used to securely store all the information needed to build the consensus mechanism and to trace the evolution of trust scores among nodes.
    Although it is orthogonal to our proposal, several proposal exist to create Blockchain solutions for IoT \cite{panarello2018blockchain,dorri2017towards,novo2018blockchain,popov2019iota}; the only requirement in our solution is the explicit support to smart contracts. 
    
\end{itemize}

Finally, in Table \ref{tab:SystemSymbols}, we report all the symbols adopted in our model.

\begin{table}
	\scriptsize
	\centering
	\begin{tabular}{||l|l||}
	\hline
		Symbol & Description \\
	 \hline \hline
		$D$ & The set of devices of the network \\
		\hline
		$BD$ & The set of basic devices, subset of $D$ \\
		\hline
		$CD$ & The set of limited devices that have sufficient computation power, subset of $D$ \\
		\hline
		$PD$ & The set of powerful devices that can train machine learning models, subset of $D$ \\
		\hline
		$d_x$ & A device that is part of $D$ \\
	    \hline
		$p_i$ & A packet sent in the network\\
		\hline
		$s^t_{\langle d_x, d_y \rangle}$ & The $interaction\_sequence$ of packets between two devices, $d_x$ and $d_y$\\
		\hline
		$C_{\langle d_x, d_y \rangle}$ & The set of $interaction\_sequence$ between $d_x$ and $d_y$ in $D$ \\
		\hline
		$f_{d_x, d_y}$ & Fingerprint model of the device $d_y$ built by the device $d_x$ \\
		&in relation to the $communication\_set$  $C_{\langle d_x, d_y \rangle}$\\
		\hline
	\end{tabular}
	\caption{Summary of the the symbols used in our model.\label{tab:SystemSymbols}}
\end{table}

\subsection{Distributed Behavioral Fingerprinting}
\label{sec:Fingerprint}

This section is devoted to the description of the adopted strategy for behavioral fingerprinting.
Our solution starts from the results described in \cite{aramini2022enhanced} and extends the proposed strategy by improving the underlying deep learning model by making it lighter, also through a tiny machine learning approach \cite{dutta2021tinyml}, and by making it more suitable for an IoT scenario.
The approach described in \cite{aramini2022enhanced} considers both network parameters, as done also in \cite{Thien2019,Jawad2019,Kyle2019}, and introduces important new features related to the packet payload.
The analysis of payload-based features is fundamental to make the behavioral fingerprint model robust also against attacks directly targeted on the surrounding Cyber Physical System. In this case, the objective of the attacker is to leave the general interaction behavior of the victim object unaltered and to modify only the content of the exchanged control packages to force the other entities in the system to adopt specific countermeasures \cite{alguliyev2018cyber}.

In particular, the approach of \cite{aramini2022enhanced} starts from the results of \cite{Thien2019} and considers the same set of networking-related features directly extracted from interaction packets among IoT devices.
Moreover, it adds two important features related to the packets' payload. 
After that, an $interaction\_sequence$ , say $is^t_{\langle d_x, d_y \rangle}$, is converted into a corresponding sequence of symbols, say $ss^t_{\langle d_x, d_y \rangle}=\{s_1, s_2, \dots, s_n\}$, obtained on the basis of the value of the aforementioned features for each packet.
The considered feature list is as follow.

\begin{itemize}
\item\textbf{Source Port Type}. The possible values are: user, system or dynamic. This feature can be converted by mapping its values to the numbers $0$, $1$, and $2$.
\item \textbf{TCP Flags}. For this feature, the original numerical values for the considered packet is preserved.
\item \textbf{Encapsulated protocol types}.  Also in this case, we can use the original numerical values already available in each packet.
\item \textbf{Interval Arrival Time (IAT)}. This feature represents the time elapsed between two consecutive packets in an $interaction\_sequence$. Therefore, given an $interaction\_sequence$, we applied a binning transformation based on the distribution of the IAT values of the involved packets and we considered $3$ indexed equal-width quantiles. In this way, each IAT value is converted into the index of the corresponding quantile.
\item \textbf{Packet Length}. The engineering of this feature considers the length of all the packets involved in an $interaction\_sequence$ and computes the corresponding frequency distribution. At this point, the first $9$ most frequent values can be mapped into $9$ bins and all the other (less frequent) values can be mapped to a single final bin.
\item \textbf{Payload Value}. This feature depends on the specific type of payload included in a packet. In particular, two macro-categories of payload can be considered, namely: categorical, and numerical. 
As for the former, the categorical payload values can be mapped to a corresponding number ranging in the interval $[0, n]$, where $n$ is the number of the possible distinct categorical values for the payload. 
Concerning the latter, a binning-based strategy can be applied. In particular, continuous payload values can be mapped to $3$ bins. The bins are identified based on the traffic generated during a controlled ``safe'' period (see Section \ref{sec:SecurityModel} for the details). Specifically, all the payload values produced in this ``safe'' period can be mapped to a central bin. At this point, all the values lower than the lower bound of such a central bin will be assigned to the first bin, and all the values higher than the upper bound of the central bin will be assigned to the last bin. 
\item \textbf{Payload Value Shift}. This feature encodes the information related to the ``variation'' in the payload values for consecutive packets. In particular, it is equal to the absolute difference between the current payload value for a packet and the payload value of the preceding packet in an $interaction\_sequence$.
\end{itemize}

A symbol, say $s_i$, is univocally associated with a combination of feature values. Packets with the same values of the involved features will be associated with the same symbol.
An example of this mapping is reported in Table \ref{tab:MappingSymbols}.

\begin{table}
	\scriptsize
	\centering
	\begin{tabular}{||c|c|c|c|c|c|c|c||}
    \hline
	Source Port type & Packet Length & TCP flag & Protocol Type & IAT & Payload Value & Payload Value Shift & symbol \\
    \hline \hline
    2 & 4 & 2 & 6 & 0 & - & - & 0 \\
    \hline
    2 & 0 & 16 & 6 & 0 & - & - & 1 \\
    \hline
    2 & 6 & 24 & 6 & 0 & 0 & 1 & 2 \\
    \hline
    2 & 0 & 17 & 6 & 0 & - & - & 3 \\
    \hline
    2 & 5 & 0 & 17 & 0 & - & - & 4 \\
    \hline
	\end{tabular}
	\caption{Example of symbol mapping\label{tab:MappingSymbols}}
\end{table}

At this point, a behavioral fingerprint solution can be seen as machine learning model trained to predict the next possible and admissible symbol in an $interaction\_sequence$.
The ratio underlying this definition is that, practically speaking, learning the behavior of an IoT node implies being able to decide in advance the next most probable action that it will carry out.

For our solution, we started from the results described in \cite{Thien2019} and in \cite{aramini2022enhanced}, in which also the features based on the payload are tested.
In both these works, the behavioral fingerprint model has been built as a Gated Recurrent Unit (GRU) neural network composed by $3$ neurons and a dense output layer.
The size of the output layer is tailored on a specific $communication\_set$, i.e., all the $interaction\_sequences$ between two IoT nodes; indeed it depends on the actual number of distinct symbols present in the overall $communication\_set$.
The experiments performed in \cite{Thien2019} proved that this approach can reach very satisfactory results by considering a sequence of $20$ symbols to be able to predict the next one.
Similarly, also the approach of \cite{aramini2022enhanced}, in spite of the addition of payload related features, obtained important results using again a window of $20$ symbols with a light training (less than $10$ epochs with a training set of about $5$K samples) to obtain an accuracy higher than $80\%$.
Both the above approaches assume the presence of a ``safe'' period in which no IoT device is corrupted. This is a fundamental requirement to train the behavior fingerprint model.

In our scenario, we are considering a more general IoT context in which also legacy devices are available (i.e., devices with medium to low computational capability).
To successfully apply such a solution to our context, it is fundamental to reduce to the minimum possible the complexity of the machine learning model in such a way to directly involve the maximum possible number of nodes.
Therefore, starting from the two solutions above, we tried to lighten the architecture as much as possible so that it could also be used by devices characterized by medium-to-low computational power. 
In particular, the solution described in \cite{Thien2019} estimates the probability of the next packet; in our case, we reduced the problem to a classification task and we just focus on the prediction of the presence or absence of a packet as the next element of an $interaction\_sequence$.
The presented neural network is again a GRU network in which we lighten up the model by cutting two of the three GRU layers and shortening the sequence of symbols required as input from $20$ to $10$.
Interestingly, this design modification allows the achievement of pretty satisfactory performance with variations with respect to the state-of-the-art solutions of $1\%$ accuracy at maximum (see Section \ref{sec:Experiments} for all the details).
Moreover, motivated by the recent introduction of {\em Tiny Machine Learning} approaches \cite{david2021tensorflow}, we proceeded by converting our model into a tiny one using the TensorFlow Lite library \cite{TensorLite} passing from a model requiring $1.6$ MB to be stored, to a model requiring only $415$ KB.
All the experiments devoted to prove the quality of the obtained results, as well as the study on the execution time for different device types, are reported in Section \ref{sec:Experiments}.

Finally, as demonstrate again in \cite{Thien2019} and \cite{aramini2022enhanced}, behavioral fingerprint models can be leveraged for anomaly detection in IoT.
Indeed, given an $interaction\_sequence$ and a sliding windows $SW$ of the last $k$ consecutive packets, the anomaly detection strategy consists in the use of the fingerprint model to predict the expected packets and to compare these results with the actual content of $SW$.
We define the misprediction rated $m\_r$ as the number of mispredictions over the total number of packets inside a window.
An anomaly is reported if the number of mispredictions observed in $SW$ is higher than a fixed threshold (typically set to $50\%$ of the packets in the sliding window).

\subsection{Distributed Consensus Mechanism to Estimate Object Reliability}
\label{sec:Consensus}

This section is devoted to describe our reliability model and the underlying distributed consensus mechanism.
In particular, as already stated above, one of the objective of our solution is to provide IoT devices with a strategy to evaluate whether to instantiate a new connection with another object based on its current behavior. 
Due to the fully distributed nature of our approach, this solution leverages an ad-hoc consensus mechanism based on the concept of word-of-mouth between devices.

Consider a scenario in which a source device $d_s$ may want to establish a new connection with an unknown target device $d_x$.
Our mechanism allows $d_s$ to obtain information about the behavior of $d_x$ from the community of nodes belonging to its neighborhood.
To do so, we define the concept of path toward a target node as an acyclic sequence of IoT nodes $pth^i_{(d_s, d_x)}=\langle d_y, d_w,\cdots,d_e \rangle$ that must be contacted to reach an {\em evaluator}, say $d_e$, which owns a behavioral fingerprint model of $d_x$ and can, hence, assess its trustworthiness.
Of course, multiple {\em evaluators} could exist for a given target node, hence we define the set of {\em evaluators} for a target node $d_x$ as $E_{d_x}=\{d_e | d_e \in D \wedge \exists f_{d_e, d_x}\}$.
As a consequence, multiple paths can exist from a source to a target node; therefore, once again, we define the set of {\em evaluation paths} as $e\_paths_{(d_s, d_x)}=\{ \langle d_y, d_w,\cdots,d_e \rangle \ |\ d_x, d_w, \cdots, d_e \ \in \ D \ \wedge \ d_e \in E_{d_x}\}$.

\begin{figure}[ht]
	\centerline{
        \includegraphics[scale=0.4]{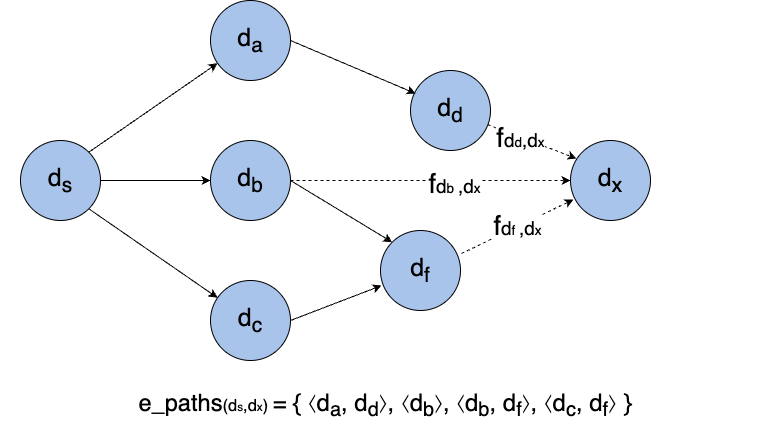}
    }
    \caption{An example of e\_paths identification in our scenario \label{fig:EPaths}}
\end{figure}

The strategy adopted by $d_s$ to obtain information from potential evaluators of the target $d_x$ is as follows.
First, it selects a $maximum\_depth$ value, which specifies the coverage range of the network. In particular, it indicates the maximum distance of propagation, in terms of number of network hops, of its request (e.g., if maximum depth is $1$ only the direct neighbors will be contacted, if it is equal to $2$ the direct neighbors will propagate this request to their neighbors, and so forth). After this it will send a request packet to all its neighbors specifying the desired target $d_x$. 
At this point, the receiving nodes will verify the $maximum\_depth$ value and they will decrease it by one. If this value is greater than zero, they will continue by adding their identifier to the packet and forwarding this request to their neighbors, thus iterating this step.
At each iteration, if the set of receiving nodes will contain an {\em evaluator} a new $path$ will be created.
This concept is illustrated in the example of Figure \ref{fig:EPaths}.

In this example, the source node $d_s$ asks its neighbors, nodes $d_a$, $d_b$, and $d_c$, information about $d_x$ by specifying a $maximum\_depth$ of $2$.
During the first iteration $d_a$ and $d_b$ decrease the $maximum\_depth$ to $1$ and propagate the packet to their neighbors (nodes $d_d$ and $d_f$). 
Node $d_b$, instead, owns a behavioral fingerprint model of $d_x$ and, therefore, performs two actions: {\em (i)} it replies to $d_s$, thus creating the path $\langle d_b \rangle$, {\em (ii)} it decreases the $maximum\_depth$ and propagates the packet to its neighbors ($d_f$).
At the second iteration, nodes $d_d$ and $d_f$ reply to $d_a$, $d_b$, and $d_c$ with the information about their models towards $d_x$, thus creating three paths, namely: $\langle d_a, d_d \rangle$, $\langle d_b, d_f \rangle$, and $\langle d_c, d_f \rangle$.
At this point $maximum\_depth$ is equal to zero and no further propagation of the original request is performed.

To inhibit any attacker from forging fake paths, all the nodes involved will add a verifiable nonce which is univocally linked to them.
To do so, we identify a solution adopting a trap-door function. 
In particular, when joining our system each node computes a hash-chain of size $q$ starting from a secret $seed$ through a cryptographic hash function $chf$.
The last value of the chain, $chf^q(seed)$, is hence made publicly available to all the other nodes (through the underlying Blockchain, see all the details below).
Every time a node is involved in a new path, it will add the next (in reverse order) element of this chain. 
Of course, the property of the hash-chaining implies that $chf(chf^{q-1}(seed))=chf^q(seed)$ for every $q$, thus providing a verification of the validity of the identifier.
As a final step, as will be clearer later, the value $chf^{q-1}(seed)$ will be made publicly available to all the nodes through the underlying Blockchain, once used in a path.

At this point, given a path $pth^i_{(d_s, d_x)} \in e\_paths_{(d_s, d_x)}$, the evaluator will reply with the estimation of the trust score of the target obtained through its behavioral model.
In our case, a trust score is defined as the complement of the average misprediction rate (see Section \ref{sub:LightModel}) obtained by the model during the $interaction\_sequences$ between the {\em evaluator} and the target node.

$$
    \tau_{pth^i} = 1 - \overline{m\_r}
$$

Where $\overline{m\_r}$ is the average misprediction rate during the most $k$ recent $interaction\_sequences$.

As shown in the example above, multiple paths can be identified from a source to a target. Each path will end with a behavioral model capable of measuring a trust score for the target.
However, different conditions, such as the obsolescence of a model, a change in the state of the target, faults in the evaluator node, could lead to wrong estimations of such a score.
To reduce the impact of such anomalies, our approach adopts a consensus mechanism based on the majority of voters. 
In particular, we impose that, in order to properly estimate the trustworthiness of a node, at least $c+1$ paths have to return values in agreement.
In our case, we consider in agreement two scores, say $s_1$ and $s_2$, such that $|s_1 - s_2| \leq tol$; where $tol$ is a suitable tolerance threshold.
It is worth observing that, as will be clearer in our Security Analysis (Section \ref{sec:SecurityModel}), given a neighbor inside the considered IoT, $c$ is identified as the maximum theoretical number of paths that an attacker can control in such a specific neighbor.
We call the set of paths returning values in agreement as {\em consensus set} $C\_S_{(d_s, d_x)}$.
Let $T_{d_s, d_x}$ be the average of all the trust scores of the $C\_S_{(d_s, d_x)}$, this value, also referred as {\em trustworthiness} value in the following, can be used by $d_s$ to decide whether to activate a communication towards $d_x$.

As an additional functionality, our approach estimates the quality of the contribution of each node involved in the estimation of $T_{d_s, d_x}$.
Indeed, this average score could be used also to compute a reliability score for the participants to the different paths in 
$e\_paths_{(d_s, d_x)}$.
In particular, the nodes involved in paths of the $C\_S_{(d_s, d_x)}$ will receive a positive feedback, whereas the members of paths returning trust scores not in agreement with the majority set will receive a negative feedback.
The extent of the negative feedback will be directly related to the bias between the returned trust score and average score $T_{d_s, d_x}$.
Our solution is designed so that a positive feedback can balance a negative one, therefore if a node is involved in two paths, one belonging to the $C\_S_{(d_s, d_x)}$ and the other with a bias in the trust score with the respect to the average, then its reliability will not have a too negative impact.
The ratio underlying this choice is related to the following reasoning. A non-attacker node can be involved in paths in which there exists also an attacker and, therefore, could receive a negative feedback; however, the same node will probably be involved also in paths belonging to $C\_S_{(d_s, d_x)}$. An attacker, on the other hand, will mostly be involved in paths with bias in the trust scores, possibly caused by its actions.
In this way, while the balance mechanism will prevent the reliability of honest nodes to go down, this effect should not happen for attackers.
At this point, we define the set $\overline{C\_S}_{(d_s, d_x)}=e\_paths_{(d_s, d_x)} \smallsetminus C\_S_{(d_s, d_x)}$ of paths with trust scores not in agreement with the {\em consensus set}.
Moreover, let $\Gamma(C\_S_{(d_s, d_x)}, d_y)$ (resp. $\Gamma(\overline{C\_S}_{(d_s, d_x)}, d_y)$) be a function returning the paths of $C\_S_{(d_s, d_x)}$ (resp. $\overline{C\_S}_{(d_s, d_x)}$) involving the node $d_y$.
More formally:

$$\Gamma(C\_S_{(d_s, d_x)}, d_y) =\{pth_i | d_y \in pth_i \wedge pth_i \in C\_S_{(d_s, d_x)}\}$$

The variation of the reliability for a node can be estimated as follows:

$$\Delta\_RL_{d_y}=\sum_{pth^i \in \Gamma(\overline{C\_S}_{(d_s, d_x)}, d_y)} \left(max(\text{-}1, \text{-}|\tau_{pth^i}-T_{(d_s, d_x)}|)\right) + \gamma \cdot |\Gamma(C\_S_{(d_s, d_x)}, d_y)|$$

Here, $max$ is a function returning the maximum value between the inputs, $\gamma$ is a parameter used to tune the impact of a positive contribution with respect to a negative one. 
In our scenario, all the nodes involved will start with a default reliability value, say $r$.
Given a node $d_y$, its reliability value will be decreased of the value $\Delta\_RL_{d_y}$, if $\Delta\_RL_{d_y}$ is less than zero; no updates at its reliability will be done, otherwise.

Actually, as described in Section \ref{sec:Model}, our approach leverages a Blockchain-based solution to trace, in a fully distributed fashion, the evolution of the behavior of objects when interacting with each other.
In our application scenario, we consider a managed Blockchain supporting smart contracts.
Our solution is based on a smart contract, say $SM$ deployed on the Blockchain, which gathers transactions from IoT nodes.
In the solution above, once the paths have been identified and the source node has received all the trust scores, it creates a transaction towards the Blockchain reporting information about all the paths, the identifier and the verifiable nonce of the involved nodes and the corresponding trust scores.
At this point, $SM$ will be executed to perform the following tasks:
\begin{enumerate}
    \item verify the nonce for each involved node \footnote{Observe that, a node can be involved in multiple paths. In this case, it will use different values of its inverse hash-chain. $SM$ will verify the consistency of all the values for a single node by linking them to the previous published element of the chain.};
    \item identify the {\em consensus set}, if it exists;
    \item compute the average trust score;
    \item compute $\Delta\_RL$ for each node;
    \item update reliability values for the nodes with $\Delta\_RL < 0$;
    \item publish a transaction reporting the results of the previous steps \footnote{The transaction will also include the last verified nonce for each involved node}.
\end{enumerate}

The information available in the Blockchain through $SM$ are, then, used by IoT nodes to identify corrupted nodes that should not be involved in the next interactions.
In particular, we assume that all the nodes with a reliability lower than a control threshold $c_{th}$ will not be engaged in future actions.

Algorithm \ref{alg:consensus} summarizes the steps of our consensus mechanism for the object reliability assessment. Observe that {\em evalPaths} is a recursive function used to compute all the $e\_paths$ between the source and target node in the network.

\SetKwComment{Comment}{/* }{ */}
\begin{algorithm}
\SetAlgoLined
\caption{Consensus mechanism for object reliability}\label{alg:consensus}

\KwData{$d_s$, $d_x$, $d_e$ \Comment*[r]{source, target and evaluator node}}
 $pth^i_{(d_s, d_x)} \gets \langle d_y, d_w,\cdots,d_e\rangle$ 
  \Comment*[r]{path toward $d_e$}
 $E_{d_x} \gets \{d_e | d_e \in D \wedge \exists f_{d_e, d_x}\}$
 \Comment*[r]{set of evaluators for $d_x$}
  $e\_paths_{(d_s, d_x)} \gets \{ \langle d_y, d_w,\cdots,d_e \rangle \ |\ d_x, d_w, \cdots, d_e \ \in \ D \ \wedge \ d_e \in E_{d_x}\}$
 \Comment*[r]{evaluation paths}
 \KwResult{$T_{d_s, d_x} \gets d_x$ reliability}
\SetKwProg{myproc}{evalPaths($d_s$, $maximum\_depth$, $d_x$)}{ is}{end}
 \myproc{}{
  $pth^i_{(d_s, d_x)} \gets pth^i_{(d_s, d_x)} + d_i$\;
  $maximum\_depth \gets maximum\_depth -1$\;
  \eIf{$maximum\_depth \gets 0 \lor d_s \in E_{d_x}$}
  { 
  \KwRet $pth^i_{(d_s, d_x)}$ \;}
  {
  \While{$d_i \in Neigh_s$}{
      $d_s$ sends a request to $d_i$ for $d_x$\;
      $e\_paths_{(d_s, d_x)} \gets e\_paths_{(d_s, d_x)} + evalPaths(d_i, maximum\_depth, d_x)$\;
   }
   }
 }
 $d_s$ selects $maximum\_depth > 0$\;
 $d_s$ computes $e\_paths_{(d_s, d_x)} \gets evalPaths(d_s, maximum\_depth, d_x)$\;
 \While{$pth^i_{(d_s, d_x)} \in e\_paths_{(d_s, d_x)}$}{
     $d_s$ requests to $d_e$: $\tau_{pth^i}$\;
 }
 $d_s$ creates a Blockchain transaction with all the information of nodes $ \in E_{d_x}$\;
\end{algorithm}

\subsection{Community-oriented Secure Delegation}
\label{sec:Delegation}

As stated above, our application scenario embraces a situation in which heterogeneous devices, belonging to the three categories described in Section \ref{sub:GeneralOverview}, collaborate to build our secure interaction scheme.
To achieve this objective, we also propose a secure delegation mechanism according to which capable devices (belonging to the $CD$ category) can delegate the training of behavioral fingerprint models to devices of the $PD$ category. Similarly, nodes of the $BD$ category can again delegate powerful devices (belonging to the $PD$ category) to train their models, and can leverage both $CD$ and $PD$ nodes for the model inference.

In particular, our approach leverages a combination of the Blockchain solution described above and an IPFS-based \cite{benet2014ipfs} strategy to exchange information about training data and the obtained models.
IPFS (InterPlanetary File System) is a peer-to-peer fully distributed file system that is typically exploited in combination with the Blockchain technology to enable secure data exchange.

At this point, our secure delegation mechanism works as follows.
Given a node $d_x$, the activation of a secure delegation starts by identifying, among the neighbors of $d_x$, reliable nodes to be delegated. The delegation can concern two aspects, namely: {\em (1)} the training of a behavioral fingerprint model; {\em (ii)} a trained model inference.
As for the former, only nodes of $PD$ can be involved. The latter instead can be demanded to both nodes from $PD$ and $CD$.
In any case, to identify reliable nodes, $d_x$ can leverage the information from the smart contract $SM$ defined in Section \ref{sec:Consensus} to obtain, for each node in its neighborhood $Neigh_{d_x}(i)$ (at any level $i$), the corresponding reliability values $RL$. 
At this point, $d_x$ will send a {\em delegation request} to all the nodes having a reliability higher than the control threshold $c_{th}$ (see Section \ref{sec:Consensus}).
Depending on their current status (including battery condition, traffic overhead, and so forth) each neighbor will decide whether to accept the request from $d_x$ or not.
In the positive case in which at least one node, say $d_n$, accepted the {\em delegation request}, our solution will proceed with the following steps.

\paragraph{Secure delegation: training a behavioral fingerprint model.}
The node $d_x$ collects the training data of its target and derive the corresponding symbols using the feature engineering task described in Section \ref{sec:Fingerprint}.
To make our solution robust to privacy issues related to the knowledge included in the symbols, we impose that $d_x$ applies a salt-based cryptographic hash function to each symbol,  $hs_i = chf(s_i, salt)$.
The choice of the adopted cryptographic hash function depends on the trade off between the need of privacy protection and computational effort to obtain the hashed symbols.
The hashed dataset will then be uploaded into a folder of IPFS and the reference of the address of such a folder will be sent to $d_n$.
At this point, $d_n$ will proceed by training a behavioral fingerprint model using the data from $d_x$ \footnote{Observe that, again for privacy reasons, the identifier of the target node of the model is not available to $d_n$.}.
Finally, if the secure delegation concerns only the training phase, $d_n$ will upload the trained model to IPFS and will share the position with $d_x$.
Instead, if the delegation concerns also the model inference, $d_n$ could retain the trained model in its memory to support $d_x$ during the subsequent model inference phase.

\paragraph{Secure delegation: model inference.}
Given a trained model, the node $d_x$ will gather the data (set of symbols) related to a sliding window, according to the approach described in Section \ref{sec:Fingerprint}.
For each symbols $s_i$ of this sliding window, $d_x$ will compute the corresponding hashed version using the salt-based cryptographic hash function $hs_i=chf(s_i, salt)$.
The obtained values will be used as input to the trained model.
At this point, two situations may occur: either $d_x$ can directly perform a model inference, or $d_x$ will share this data with its delegate $d_n$.
In both cases, the output will be the misprediction rate measured in the corresponding sliding window.

\begin{figure}[ht]
	\centerline{
        \includegraphics[scale=0.4]{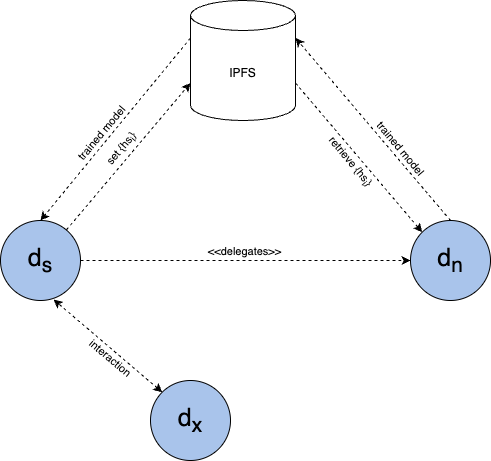}
    }
    \caption{An example of our secure delegation strategy \label{fig:delegation}}
\end{figure}

Figure \ref{fig:delegation} sketches the solution described above and the use of IPFS to exchange both the training data and the trained behavioral fingerprint model.

Algorithm \ref{alg:delegation} summarizes the steps of our community-oriented secure delegation mechanism.

\SetKwComment{Comment}{/* }{ */}
\begin{algorithm}
\SetAlgoLined
\caption{Secure Delegation mechanism}\label{alg:delegation}

\KwData{$d_x$, $d_t$, $c_{th}$ \Comment*[r]{source node, target node, control threshold}}
 $hs_i \gets chf(s_i, salt)$ \Comment*[r]{salt-based cryptographic hash function for symbol $s_i$}
\KwResult{$\Delta\_RL_{d_t}$ 
\Comment*[r]{$d_t$ misprediction rate}}
 $d_x$ gets from the Blockchain the list of nodes $Neigh_{d_x}(i)$ \;
 \While{$d_n \in Neigh_{d_x}(i)$}{
    \If{$RL_{d_n} > c_{th}$}{
        $d_x$ sends a delegation request to $d_n$\;
        \If{$d_n$ status $\gets$ ok}
            {$d_n$ accepts $d_x$ request\;
            $d_x$ collects training data of $d_t$\;
             $d_x$ applies $hs_i \gets chf(s_i, salt)$\;
             \eIf{secure delegation includes only the training phase}
                {$d_n$ trains a behavioral fingerprint model using $d_x$ data\;
                $d_n$ uploads the trained model to IPFS\;
                $d_n$ shares the position with $d_x$\;}
                {$d_n$ (or $d_x$) computes $\Delta\_RL_{d_t}$\;}
            }
         }
 }

\end{algorithm}

\section{Security Model}
\label{sec:SecurityModel}

This section is devoted to the security model underlying our solution.
In the next sections, we introduce both the attack model and the security analysis proving that our approach works also in presence of attacks.
Since the main contribution of our strategy is the construction of a mechanism to evaluate the trustworthiness of IoT nodes, in our security analysis we consider classical attacks to reputation systems such as those mentioned in \cite{corradini2022two,HeBuCh15}.

\subsection{Attack Model}
\label{sub:Attack-Model}

Preliminary, observe that our solution in the stationary scenario considers enough nodes available to carry out the steps required by our approach.
As a consequence, we do not focus on initial stages possibly characterized by anomalous situations in which the IoT network is not yet active or complete.

With that said, before analyzing the security properties of our model, we outline the following starting assumptions:

\begin{enumerate} [label=\textbf{A.\arabic*}]

\item \label{listEl:collude} An attacker can control at most $c$ paths of a set $e\_paths_{d_s,d_x}$, between any pair of nodes, $d_s$ and $d_x$.



\item \label{listEl:safe} There exists a safe stage in which behavioral fingerprint models can be computed in the absence of attacks.

\item \label{listEl:control} An attacker cannot control all the behavioral fingerprint models associated with IoT nodes.

\item \label{listEl:background} An attacker has no additional knowledge derived from any direct physical access to IoT objects (especially about the paths linking nodes).

\item \label{listEl:blch} The Blockchain technology exploited to implement the support public ledger is compliant with the standard security requirements already adopted for common Blockchain applications.

\item \label{listEl:seed} An attacker cannot have access to the secrets to generate the hash chains of IoT nodes.

\item \label{listEl:hash} The adopted cryptographic hash function is robust again collision, preimage and second preimage attacks.

\end{enumerate}

As stated above, our model ensure a list of security properties (SP, in the following), which are listed below.

\begin{enumerate}[label=\textbf{SP.\arabic*}]
	\item \label{blck} Resistance to attacks to the Blockchain and the smart contract technology.
    \item \label{SPA} Resistance to Self-promoting Attacks.
    \item \label{WA} Resistance to Whitewashing or Self-serving Attacks.
    \item \label{SLA} Resistance to Slandering or Bad-mouthing Attacks.
    \item \label{OPP} Resistance to Opportunistic Service Attacks.
    \item \label{BSA} Resistance to Ballot Stuffing Attacks.
    \item \label{DoS} Resistance to Denial of Service (DoS) Attacks.

\end{enumerate}

\subsection{Security Analysis}
\label{sub:Security-Analysis}

This section is devoted to the analysis of the security properties presented above to prove that our approach can ensure them.
In the next sub-sections, we analyze each of these properties in details.

\subsubsection{\ref{blck} -  Resistance to attacks to the Blockchain and the smart contract technology}

This category of attacks aims at finding vulnerabilities in the Blockchain and the smart contract technology adopted in our approach.
This technology has been subject of studies from the research community in the recent years. Anyway, the security of Blockchain is still under the spotlight and represents an open issue \cite{kushwaha2022systematic,singh2021blockchain,bhutta2021survey,idrees2021security}.
The approach presented in this paper is not devoted on facing security challenges on Blockchain, but focuses on its application as a secure public ledger to support a collaborative fully distributed approach for secure device interactions in IoT.
Therefore, we assume that the Blockchain and the smart contract technology can be considered secure (Assumption \ref{listEl:blch}).
Our approach is, hence, orthogonal to any solution, available in the scientific literature, to improve the security of Blockchain that can, hence, be applied in our context to guarantee our assumption.

\subsubsection{\ref{SPA} - Resistance to Self-promoting Attacks}
\label{subsub:SP1}

In this case an attacker controlling a node or a set of nodes could work to manipulate the reliability and trustworthiness of a target node.
This attack can be done by either a single node or through a joint actions of different colluding nodes.
Concerning the first case, actually, a node cannot alter the perception that other nodes have of it, because the trustworthiness of a node is estimated using behavioral fingerprinting. Therefore, any variation in its typical behavior would increase the misprediction of the models associated with this node.
To alter this value, the attacker should have access to the nodes holding a fingerprint model towards the target. However, this cannot happen thanks to Assumptions \ref{listEl:safe}, \ref{listEl:control}, and \ref{listEl:background}.
Also the reliability cannot be altered because it is associated with the collaborative approach described in Section \ref{sec:Consensus}.
As such, the attacker should include its node always in paths belonging to {\em consensus sets} to avoid the detriment of its reliability.
However, for Assumption \ref{listEl:background} an attacker cannot know the topology of the network nor force the inclusion of a node into a path due to the security mechanism leveraging the cryptographic hash chaining strategy described in Section \ref{sec:Consensus}, which thanks to Assumptions \ref{listEl:seed} and \ref{listEl:hash} cannot be broken or forged.

According to the second strategy, an attacker could leverage different nodes colluding to positively change the trustworthiness and reliability of a node.
As for the trustworthiness, once again, due to Assumptions \ref{listEl:safe}, \ref{listEl:control}, and \ref{listEl:background} the attacker cannot control all the behavioral fingerprint models; moreover, she/he cannot forge a trustworthiness score for the target node by leveraging a collaborative self-promoting attack. Indeed, thanks to Assumption \ref{listEl:collude}, the attacker can only control $c$ paths in the set of $e\_paths$ between the target node and any other node in the system. However, the trustworthiness value of a node can only be estimated if there exist at least $c+1$ paths with a trust score in agreement. Therefore, also thanks to Assumptions \ref{listEl:seed} and \ref{listEl:hash}, this attack cannot happen.
As for the reliability, as already seen for the single node version of this attack, because the estimation of the reliability derives from the consensus mechanism and since the attacker cannot forge paths nor control more than $c$ real paths in an $e\_paths$ set (Assumptions \ref{listEl:seed}, \ref{listEl:hash}, and \ref{listEl:collude}), this attack cannot be carried out. 

Finally, in any case the attacker cannot alter the computed trustworthiness and reliability scores from the Blockchain thanks to Assumption \ref{listEl:blch} and the security property \ref{blck}.

\subsubsection{\ref{WA} - Resistance to Whitewashing or Self-serving Attacks}

This attack concerns any attempt of a malicious node to clean its trustworthiness and reliability scores and to be involved again in the activity of the network.
In our approach, trustworthiness scores are computed through the evaluation of the behavior of the target nodes by exploiting an existing trained fingerprint model.
Therefore, also thanks to Assumption \ref{listEl:control}, an attacked node cannot whitewash this score as it depends on the models owned by the surrounding nodes.
As for the reliability of a node when participating to the collaborative consensus-based mechanism, the corresponding value is permanently stored in the underlying Blockchain.
In the absence of attacks, the reliability is set initially to a default positive (over the control threshold) value $\gamma$, and it can only be reduced (no positive increment) if the interactions of the node during the execution of the consensus algorithm are not evaluated positively by the community.
If the reliability of a node is under a control threshold (see Section \ref{sec:Consensus}), this node will not be involved in the next activities by the other members of the IoT.
Depending on the security requirement of the considered scenario, an under-threshold reliability score can stay so for a time interval of ban, say $t_{ban}$.
After that, the reliability is restored to the default initial value $\gamma$.
Of course, for critical scenarios this value could be infinite so that the restore of a node could be done only through a manual intervention of a system administrator.

It is worth observing that, in IoT, one of the main problems is related to the difficulty of assessing a unique identifier for a device.
In our case, there is a directly relationship between a device identifier and its profile on the underlying Blockchain.
Of course, an attacker could perform a whitewashing attack on a node by exiting the system and re-introducing the device with a different (forged) identifier.
To face this situation, we adopt a pessimistic attitude approach, which imposes that newly introduced devices will be associated with a negative (under the control threshold) reliability \cite{Yu*12,GaBaSri08,LaGuSe16}. 
Under this assumption, a new device will start in a banned state (no other other node will interact with it) until its reliability is set to the default value after $t_{ban}$.
In this way, attempting a whitewashing by forging a new identifier for a device would result again on the node being banned for $t_{ban}$. Therefore, no advantage is obtained by the attacker.

\subsubsection{\ref{SLA} - Resistance to Slandering or Bad-mouthing Attacks}
\label{subsub:SLA}

According to this typology of attack, the adversary tries to ruin the trustworthiness and/or the reliability of a target node to force its exclusion from the system.
As for the trustworthiness, this can only be estimated through existing trained behavioral fingerprint models.
For the Assumption \ref{listEl:control}, the attacker cannot control all the models referring to its target.
Our approach is designed to use a consensus strategy to estimate the trustworthiness of a node by leveraging all the information obtained by the possibly different trained models describing its behavior.
When it comes to the reliability, instead, this is evaluated based of the quality of the interactions of a node during the execution of the consensus algorithm.
In this case, the adversary can attempt to execute two different strategies.
Because the overall reliability variation derived from an interaction depends on the difference between the result provided by the {\em consensus set} and that of the paths which a node is involved in, the attacker could force a wrong estimation for a path it is also involved in.
In this way, all the nodes (including the attacker) in such a path will be negatively evaluated.
Our strategy implies a balancing mechanism according to which if a node is involved in both a positive and a negative path, no negative variation is recorded for its reliability.
Our assumption here is that a non-malicious node will be involved in different paths that cannot be all controlled by an attacker (Assumption \ref{listEl:collude}).
Of course, if a node is only linked to an attacked one, it can be considered under the control of the attacker itself, and, hence, it should be excluded by the system.
According to the second strategy, instead, the attacker could try to forge false paths returning results very different from any {\em consensus set} and involving the target node.
In this way, she/he could effectively cause a detriment on the reliability of such a node.
As a countermeasure, our approach includes a mechanism to ensure the real membership of a node to a path.
Indeed, when a path is formed, each involved node would add a verifiable nonce uniquely related to it.
This solution is obtained through a cryptographic hash chain that each node build when joining the system.
The first element of the (inverse) chain is stored on the Blockchain and is used to verify the correctness of the following values.
At the end of the execution of the consensus algorithm the used elements of the chains are made public available on the Blockchain (see Section \ref{sec:Consensus}).
Of course, this implies that no one could re-use the information of a path to perform a {\em reply-attack} because the verifiable nonce must not be already available in the Blockchain (so that $chf(chf^{q-1}(seed))=chf^q(seed)$).
This strategy, also thanks to Assumptions \ref{listEl:blch}, \ref{listEl:seed}, and \ref{listEl:hash} makes our approach robust against this attack.

\subsubsection{\ref{OPP} - Resistance to Opportunistic Service Attacks}
\label{subsub:OPP}

A malicious node could selectively behaves good or bad, opportunistically.
This strategy can be carried out on both the standard interactions with the other nodes and on the interactions related to the consensus mechanism.
As for the former aspect, behavioral fingerprint models can successfully detect any change in the behavior even if selective. 
Indeed, these models are built under the Assumption \ref{listEl:safe} in which no malicious behavior is present.
As for the second strategy, the idea underlying it is that the attacker knows of the balancing effect in the computation of the reliability and tries to leverage this feature to partially attack the network still preserving its status in it.
However, because our approach enforces the existence of a {\em consensus set} to evaluate the trustworthiness of a node, thanks to Assumption \ref{listEl:collude}, no advantage can be obtained by the attacker.
Of course, she/he could selectively cause the detriment of the reliability of nodes that are only linked to the attacker (and, hence, are not involved in other {\em honest} paths). However, as stated above, if a node is reachable only through the attacker, it can be assumed under her/his control. For this reason, the consequent isolation of such a node (due to a low reliability) is actually intended in our solution.

\subsubsection{\ref{BSA} - Resistance to Ballot Stuffing Attacks}

According to this typology of attack, an adversary could exploit the position of a node in the network to positively increment the trustworthiness and/or reliability of a target (malicious) node.
As for the trustworthiness, the attacker would need to control all the behavioral fingerprint models of the target, which is forbidden by Assumption \ref{listEl:control}.
Moreover, she/he cannot control all the paths towards the target so to selectively hide the {\em honest} models and privilege the controlled ones to obtain a {\em consensus set} (Assumption \ref{listEl:collude}).
Finally, for Assumptions \ref{listEl:seed}, and \ref{listEl:hash}, she/he cannot forge artificial paths to favor the access to controlled (malicious) models.
As for the reliability, the attacker cannot positively increment it for a node because our approach records only negative variations.
Once under the control threshold, the reliability can be restored only after a ban period $t_{ban}$ (see the description for the security property \ref{WA}).
Finally, recall that, reliability scores are stored on the Blockchain, which enforces that no devices can corrupt or change such scores, either positively or negatively.

\subsubsection{\ref{DoS} - Resistance to Denial of Service (DoS) Attacks}

Denial of Services (DoS) attacks try to bog down a system by overflooding it with a very high number of (dummy) transactions.
In our approach this attack could also result in the impossibility for nodes to gather information about the trustworthiness and reliability of the other nodes.
Indeed, the goal of the attacker could be to prevent the identification of the {\em consensus set} to estimate the trustworthiness of a node.
Although it may represent a problem also in our context, our strategy does not 
directly deal with this typology of attack. However, it is important to mention that, our approach does not add any advantage to an adversary performing this typology of attack.
For this reason, any existing strategy conceived to prevent/face DoS attacks in IoT could be included in our approach, such as the solutions presented in \cite{hussain2020iot,baig2020averaged,abughazaleh2020attacks,shurman2019iot}.

In IoT, DoS attacks can take the form of a Sleep Deprivation Attack (SDA) whose objective is the power consumption of the devices to exclude them from the system through a battery drain.
As for this aspect, our approach natively supports a countermeasure.
Indeed, when performing a DoS attack a node alters its standard behavior. Such information is detectable by our behavioral fingerprint models; therefore, IoT nodes can safely discard all the requests from nodes whose behavior is anomalous, thus preventing SDA attacks to happen.

\section{Experiments}
\label{sec:Experiments}

This section is devoted to our experiments for validating the proposed approach.
In particular, in the next sub-sections we report in details the performance evaluation of our solution to build a behavioral fingerprint model, the tests to identify the best tuning configurations, the experiment devoted to assess the quality of our delegation strategy, as well as the performance of the overall approach on different type of devices, and, finally, we show the results of our solution for the anomaly detection using our consensus-based algorithm.

\subsection{The Underlying Dataset}
\label{sec:Dataset}

To conduct our experiments we started from the same dataset and approach of \cite{aramini2022enhanced}. 
In particular, we leveraged the dataset described in \cite{hamza2019detecting} and available at \url{https://iotanalytics.unsw.edu.au/attack-data}.
The dataset is composed of two parts: the raw packet traces, and the flow counters.
The data concerns the interactions of $27$ IoT nodes; among these $10$ devices were also included in attack traffic.
Benign and attack traffic has been recorded for two periods, namely from May $28$th $2018$ to June $20$th $2018$, and from October $10$th $2018$ to October $29$th $2018$.
The attack traffic was properly labeled and occurred in the periods from June $1$st $2018$ to June $8$th $2018$, on June $20$th $2018$, and from October $20$th $2018$ to October $27$th $2018$.
The information about the attacks comprises the start and end time, the flow influenced by the attack, the type of the attack, the bit-rate of the attack, the attacker identifier, and the victim identifier.

A limitation of this dataset in our context is related to the fact that the considered scenario concerns a centralized environment in which a {\em hub} collected the message exchanges. Because the central {\em hub} is not the intended recipient of such messages, the collected packet information cannot include payload data (due to the message encryption).
However, payload-based features are an important component in our approach and, therefore, we adopted the strategy proposed in \cite{aramini2022enhanced} to alter the previous dataset and include synthetic payload data.
In particular, among all the packets available in the original dataset, the ones carrying a payload can be identified by checking the {\tt PSH TCP} flag.
At this point, we used the algorithms originally proposed in \cite{aramini2022enhanced} to generate both quantitative and categorical payload data. 
Quantitative payloads generation simulates devices like temperature, pressure, or humidity sensors (Algorithm \ref{alg:1}).

\SetKwComment{Comment}{/* }{ */}

\begin{algorithm}
\caption{Algorithm for quantitative payload generation \cite{aramini2022enhanced} \label{alg:1}}
\KwData{$R$ = [lower bound, upper bound], $HOP$, $n$}
\KwResult{$PL$\Comment*[r]{PL is a list of $n$ payload values}}
$i\gets 1$\;
$PL\_0 \sim U(R)$\;
\While{$i < n$}{
$PL_i \sim$ U(R$ \cap [PL_{i-1} - HOP , PL_{i-1} + HOP])$\;
$i \gets i + 1$\;
}
\end{algorithm}

In practice, this algorithm takes in input the range values in which the generated payloads should be contained ($R$), the number of consecutive payload values that should be generated ($n$) and the maximum gap admitted between $2$ consecutive payload values ($HOP\geq 0$). 
Hence, it generates the first payload value $PL_0$ through a uniform sampling in $R$. Each successive payload value, say $PL_i$, is uniformly sampled in an interval centered around the value of the previous payload, say $PL_{i-1}$, with a size of $2 \cdot HOP$ . The algorithm controls the rate of variation of the generated quantitative payload values based on the $HOP$ parameter.

As for the categorical payload generation, we adopted the Algorithm \ref{alg:2} originally proposed, once again, in \cite{aramini2022enhanced}.

\SetKwComment{Comment}{/* }{ */}
\begin{algorithm}
\caption{Algorithm for categorical payload generation \cite{aramini2022enhanced} \label{alg:2}}
\KwData{$R=\{ val_1, \ldots, val_q \}$, $n$, $stabilityPeriod$=[min, max]}
\KwResult{$PL$\Comment*[r]{PL is a list of $n$ payload values}}
$i \gets 0$\;
$PL_0 \sim U(R)$\;
\While{$i < n$}
{$ STAB \leftarrow stabilityPeriod$\;
\eIf{$i+STAB < n$}{
	$val \sim  U(R)$\;
 	$PL_{i,...,i+STAB}\gets val$\;
	  $i \gets i+STAB+1$\;
    }{
	$val \sim  U(R)$\;
	$PL_{i,...,n} \gets val$\;
	$i \gets n$\;
    }
}
\end{algorithm}

This algorithm accepts the list of categorical values ($R$), the total number of different payloads to be generated ($n$), a stability period representing the time interval in which the categorical value should not change.
At this point, the algorithm extracts a duration for a categorical value in the range defined by the stability period. After that, it randomly extracts the corresponding value from $R$. This value will be contained in all the packets exchanged during a time window equal to the duration extracted above.
The stability period is used to control the oscillation frequency of the categorical payload.
Some statistics about the obtained dataset are reported in Table \ref{tab:dataStats}.

\begin{table}[ht]
	
	\centering
	\begin{tabular}{||l|c|c||}
		  \hline
	Type of communication &  Min $\#$ of packets & Max $\#$ of packets \\
    \hline \hline
		 Benign & 12,793 & 97,256\\
		\hline
		Benign with payload  & 4,670 & 39,000\\
		\hline
	    Malign & 6,971 & 89,148\\
		\hline
	    Malign with payload & 2,196 & 8,694\\
		\hline
	\end{tabular}
	\caption{Statistics of the dataset considered in our study.\label{tab:dataStats}}
\end{table}

\subsection{Analyzing the Performance of our Lightweight Fingerprint Model}
\label{sub:LightModel}

This section reports the details about the training and the performance obtained for our lightweight behavioral fingerprint model.
To conduct this experiment, we randomly selected the {\em communication set} of three different devices from the dataset introduced in Section \ref{sec:Dataset}.
As stated in Section \ref{sec:Fingerprint}, our solution starts from the results reported in \cite{aramini2022enhanced,Thien2019}, and we strove to build a lighter version of the models proposed in this previous study to cope with the limitation of the considered IoT context.
To prove the performance of our solution, we performed a comparative evaluation between our new model and the model of \cite{aramini2022enhanced}. The results of this comparative analysis are reported in Table \ref{tab:AccuracyModel}.

\begin{table}[ht]
	\centering
	\begin{tabular}{||c| c| c||}
    \hline
	Model & Accuracy & F-Measure\\
    \hline \hline
        Model of \cite{aramini2022enhanced}  & 79\% & 78.23\%\\
        \hline
		Our Model & 78.6\% & 79.55\%\\
		\hline
		Tiny Machine Learning Version & 78.6\% & 76\%\\
		\hline
	\end{tabular}
	\caption{Accuracy of the models on Test Set. \label{tab:AccuracyModel}}
\end{table}

From this table, we can see that, despite its simpler structure, our model is capable to mostly match the performance of \cite{aramini2022enhanced} with a maximum difference of 1\% in accuracy. 
This result is preserved also for the tiny version of our model obtained through the conversion done with TensorFlow Lite \cite{TensorLite}.

It is worth mentioning that the lightening process has allowed us to reduce the number of model parameters by $60\%$ as shown in Table \ref{tab:ParametersModel}, while maintaining the desired performance.

\begin{table}[ht]
	\centering
	\begin{tabular}{||c| c||}
    \hline
	Model & Number of Parameters\\
		\hline \hline
		Model of \cite{aramini2022enhanced}  & 260k \\
		\hline
		Our Model & 106k \\
		\hline
	\end{tabular}
	\caption{Number of Parameters. \label{tab:ParametersModel}}
\end{table}

\subsection{Anomaly Detection: Window Size Selection}

After training the model for the prediction of the next symbol, it is possible to build our solution for anomaly detection to identify changes from the normal behaviour of the analyzed object. 

\begin{figure}[ht]
	\centerline{
        \includegraphics[scale=0.5]{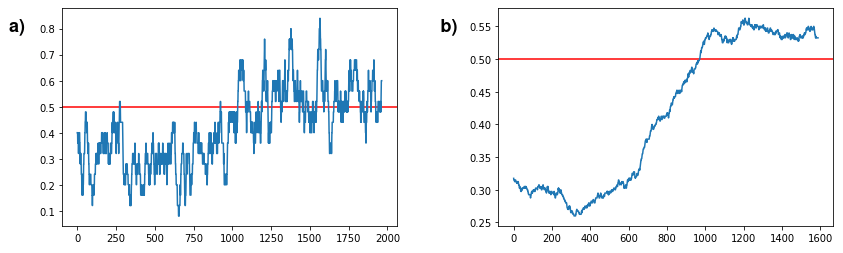}
    }
    \caption{Traffic analysis with windows of different sizes: \emph{a)} 25 packets size and \emph{b)} 400 packets size.}
    \label{fig:ComparisonSizes}
\end{figure}

To do so, we check the misprediction rate of the next symbol in a given window of consecutive packets. In our approach, we detect an anomaly when more than half of the packets predicted are different form the packets received. 
Following the reasoning above, our anomaly detection strategy strongly depends on the correct size of the chosen window. 
Therefore, using the models obtained from the previous experiment, we proceeded by analyzing the misdirection rate of the devices using windows of different sizes. 
In particular, in our experiment we considered an {\em interaction\_sequence} of $2,000$ packets, in which the first half part of the sequence represents benign traffic and the second half malign one.
We tested our solution with different window sizes and, for each of them, we analyzed the misprediction rate and, in particular, we focused on the difference between the maximum and minimum peaks of this curve. As a result, we obtained that the bigger the window size the more stable the obtained curve. 
In Figure \ref{fig:ComparisonSizes}, we report the result of this analysis for both a window size of $25$ packets and one of $400$ packets.
In this figure, the x-axis reports the sliding window number, while the y-axis indicates the misprediction rate. The anomaly threshold is fixed to $0.5$ (meaning that an anomaly is detected if half of the packets inside a window are incorrectly predicted).

As we can see from this figure, the size of the window drastically changes the oscillation of the misprediction rate curve. This oscillatory behavior can, of course, lead to a not stable anomaly detection.

\begin{figure}[ht]
	\centerline{
        \includegraphics[scale=0.4]{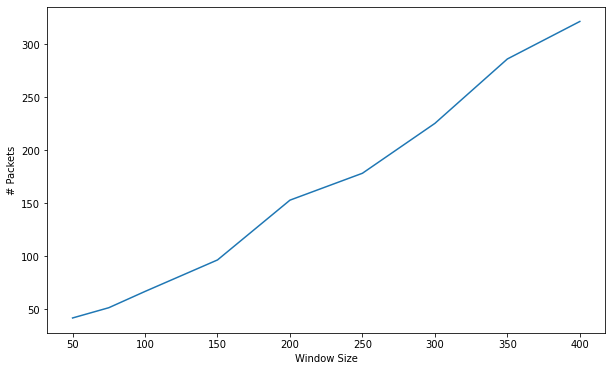}
    }
    \caption{Number of Packets required to detect an anomaly.}
    \label{fig:NpacketSizes}
\end{figure}

Intuitively, bigger window sizes would result in a greater number of packets to detect an anomaly.
In particular, from Figure \ref{fig:NpacketSizes}, we can see how the number of required packets to detect an anomalous behavior is directly proportional to the size of the window.

Therefore, to select the best window size, starting from the difference of the maximum and minimum peaks of the misprediction curve, we leverage the \emph{Kneedle} algorithm \cite{satopaa2011finding}. Specifically, this algorithm tries to find the elbow/knee in a curve by selecting the {\em right} operating point for a given system. 

\begin{figure}[ht]
	\centerline{
        \includegraphics[scale=0.4]{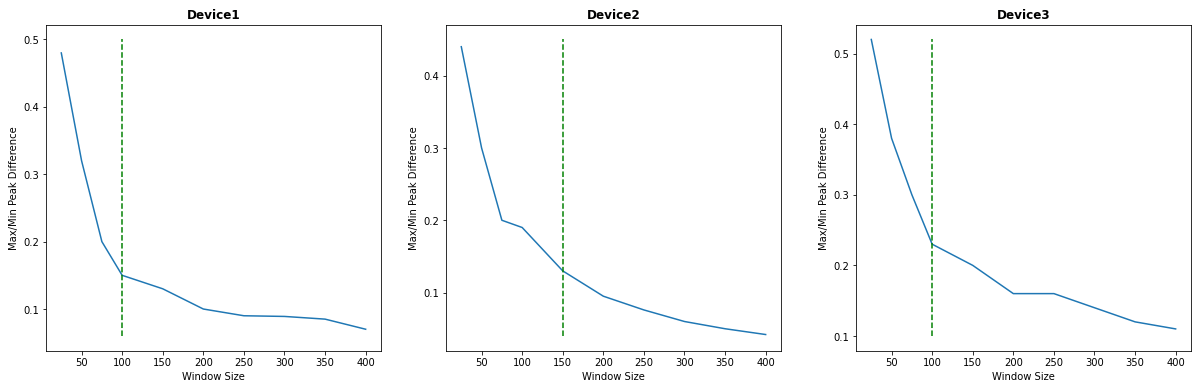}
    }
    \caption{Difference between Minimum and Maximum Peaks.}
    \label{fig:PeakWindowSizes}
\end{figure}

In Figure \ref{fig:PeakWindowSizes}, we demonstrate the application of this strategy to identify the best window size for {\em interaction\_sequences} involving three different devices.  The results show that the \emph{Kneedle} algorithm returned a correct window size of about $100$ packets for all the three devices.

\subsection{Analysis of the Execution Time for Different Types of Device}

In this section, we study the execution time required by the different tiers of devices to train and execute the fingerprint model described in Section \ref{sub:LightModel}. 
The objective of this experiment is to verify the advantages, in terms of running time, introduced by our secure delegation strategy.


In this experiment, we compare three different devices, one belonging to the class of {\em Basic Devices}, one to the class of {\em Capable Devices}, and one belonging to the {\em Powerful Device} class. 
In particular, as for the {\em Basic Device}, we used an emulation device equipped with a single core $1$GHz Arm processor (ARM $1176$) using the QEMU emulator. This kind of processor is usually employed in objects such as smart locks, smart thermostats or simple board like the first version of the Raspberry Pi family. 
Concerning the {\em Capable Device} class, we selected a general purpose single-board personal computer, namely {\em Raspberry Pi $4$}. These devices are very common examples of IoT nodes \cite{kumar2017air,kumar2016iot}. 
Finally, as for the {\em Powerful Device} class, we employed a device equipped with an eight-core Desktop CPU (Ryzen $7$ $5800$x).

We tested the performance of these devices for both training and model inference and we reported the results in Table \ref{tab:ExecutionTimes}.

\begin{table}[ht]
	\centering
	\begin{tabular}{||c| c | c| c||}
    \hline
	& Single-Core ARM & Raspberry Pi 4 & Powerful Device \\
		\hline \hline
		Training Epoch Time (seconds) & 671 & 30 & 3 \\
		\hline
		Prediction Time (milliseconds) & 1500 & 86 & 7 \\
		\hline
		TinyML Prediction model Time (milliseconds) & 0.4 & 0.4 & 0.004 \\
		\hline
	\end{tabular}
	\caption{Training and model inference execution times for different classes of device \label{tab:ExecutionTimes}}
\end{table}

From this table, we can see how the class of {\em Basic Devices} requires more than $10$ minutes to complete an epoch compared to the $3$ seconds required by the class {\em Powerful device}. 
The {\em Capable Device} class would require about $30$ seconds for each training epoch.
In our experiment, we found that the training needs about $10$ epochs to reach satisfactory performance with a training set of about $5K$ samples.
For this reason, the overall training would be mostly prohibitive for the {\em Basic Device} class. Instead, it would require a moderate energy consumption for the class of {\em Capable Devices} (about $5$ minutes of computation, on average).
The training cost would be negligible for the {\em Powerful Device} class (about $30$ seconds on average). Thus making our secure delegation strategy particularly advantageous in this case. 

As for the model inference, the times required to predict a single symbol appears sustainable in all the cases, with a maximum of $1.5$ seconds for the {\em Basic Device} class.
Interestingly, this value reduces drastically after the conversion of our model to a TinyML solution.

\subsection{Anomaly Detection using the Consensus Algorithm}

Although, thanks to the extreme lightness of our model, all the classes of devices can train a behavioral fingerprint model, either directly ({\em Powerful Devices} and {\em Capable Devices}) or through our secure delegation mechanism, and perform model inference, this can happen only in presence of a {\em safe} period that can be assumed, for example, during the {\em start-up} of the network.
For this reason, only a percentage of nodes can ultimately have behavioral models for some of their neighbors. 
To enable the propagation of trustworthiness data on the whole IoT network, we designed a distributed consensus mechanism to estimate the trust values associated with a node (see Section \ref{sec:Consensus}). 

The experiment described in this section aimed at evaluating the performance of our strategy for distributed anomaly detection. In particular, we tested the average time required by a generic node in the network to detect an anomalous behavior of a peer, estimated through our distributed consensus mechanism.
To do so, we built a simulated IoT by using device emulation solutions (such as QEMU and VirtualBox) equipped with a telemetry system (to log all packet exchange) written in Python. We implemented our solution and used the dataset described in Section \ref{sec:Dataset} to simulate real interactions among the nodes.
To perform our experiment, we injected data from a malicious node in the simulated network and collected all the information and data exchanged among the nodes.
In particular, starting from the dataset generated in Section \ref{sec:Dataset}, we crafted different communications between devices inside the network maintaining the integrity of the normal behavior for the involved nodes. The crafted communications are composed of half benign and half malign traffic in order to monitor the shift between the normal behaviour and an anomalous one. 
Specifically, we measured the number of additional packets necessary to detect an anomaly through our consensus algorithm.
We found that, this number is about $61$, on average. 
Clearly, depending on the dynamicity of the network (in terms of the frequency of packet exchange among nodes) this could correspond to very different detection times, spanning from few seconds to some minutes.
However, it is worth noting that, when a node is compromised and exhibits an anomalous behavior, direct neighbors owing a behavioral fingerprint model will detect the anomaly almost instantaneously. 
The additional $61$ packets reported above represents the average propagation effort required by our system to inform {\em any} other interested node, inside the whole IoT network, of the behavior change of a malicious actor.


\section{Conclusion}
\label{sec:Conclusion}
In the last years, we assisted to an enormous increase in number and potentialities of IoT devices. From simple sensors/actuators to smarter nodes, all these actors enable the IoT network with complex monitoring, automation, and decision-making capabilities. Obviously, in this scenario, where IoT services and applications are
intimately associated with people and are more and more autonomous, the issue of trust management becomes a major challenge. This paper gives a contribution in this setting, designing a complete framework to assess the trustworthiness of an object before contacting it. Our approach, based on collaboration and delegation, proceeds through two steps. At an initial stage of the network, behavioral models representing the expected conduct of every node are built thanks to a novel tiny machine learning algorithm suitable for limited devices. In the following fully operational state, every node is equipped with the possibility to detect possible variations or anomalies in the behaviour of other objects and then decide to contact them. This feature is provided by a distributed consensus mechanism based on the concept of word-of-mouth between neighbors. Moreover, all the nodes, even the less smart ones can participate to our framework thanks to a secure delegation mechanism, according to which they can entrust the training of behavioral fingerprint models to more powerful devices.
Furthermore, Blockchain is used to both store the
reliability and trust scores related to the behavior of objects and to identify the best peers to contact
to enable our collaborative approach.

The research directions taken in this paper can be considered as starting point, since we plan to make further investigations in this field in the next future. For instance, behavioral fingerprinting of group of objects can be analysed to detect specific typology of complex and distributed attacks in the network. A behavioral group fingerprint solution can be seen as machine learning model trained to predict the next possible distributed attack in the network.

\end{document}